%% file: user_expertise.tex
\documentclass{sig-alternate}

\usepackage[
bookmarks=false
,bookmarksnumbered=true
,hypertexnames=false
,breaklinks=true
]{hyperref}
\hypersetup{
  pdfauthor={Anonymous},
  pdftitle={},
  pdfsubject={Workshop on Query Understanding and Reformulation for Mobile and Web Search (QRUMS)'15},
  pdfkeywords={H.3 [Information Storage and Retrieval]},
  pdfcreator={LaTeX with hyperref package},
  pdfproducer={pdflatex}
}
\usepackage{comment}
\usepackage{booktabs}
\usepackage{listings}
\usepackage{paralist}
\usepackage{xspace}
\usepackage{graphicx}
\usepackage{graphics}
\usepackage[utf8]{inputenc}
\usepackage{epsfig}
\usepackage{multirow}
\usepackage{amsmath}
\usepackage{mathtools}
\usepackage{algorithm}
\usepackage{algorithmic}
\usepackage{color}

\newcounter{todocnt}

\newcounter{latercnt}
\newcommand{\old}[1][$\bullet\bullet\bullet$]{%
{\refstepcounter{latercnt}%
\textcolor{red}{$\bullet\bullet\bullet$\ \sf \scriptsize Old text (\thelatercnt): #1}}\xspace}

\usepackage[square,comma,numbers,sort&compress,sectionbib]{natbib} 
\def\:{\hskip0pt} 
\usepackage{booktabs}
\usepackage{tabularx}
\newcommand{\mypar}[1]{\medskip\noindent\textbf{#1}~}
\newcommand{\shrink}{\vspace{-1.5ex}}

\def\sharedaffiliation{
\end{tabular}
\begin{tabular}{c}}

\newfont{\mycrnotice}{ptmr8t at 7pt}
\newfont{\myconfname}{ptmri8t at 7pt}
\let\crnotice\mycrnotice%
\let\confname\myconfname%

\begin{document}
\conferenceinfo{Workshop on Query Understanding and Reformulation for Mobile and Web Search (QRUMS) }{'16, San Francisco, California, USA}

\title{Understanding Experts Search Behaviour for Serving Verified Search Results}
\title{The Impact of Technical Domain Expertise \\ on Search Behavior and Task Success}
\title{The Impact of Technical Domain Expertise \\ on Search Behavior and Task Outcome}

\newcommand{\rqmain}{How does technical domain expertise influence search behavior?}
\newcommand{\rqone}{\textbf{RQ1:}\textsl{How to define user expertise?}}
\newcommand{\rqtwo}{\textbf{RQ1:}\textsl{How to define experts topics?}}
\newcommand{\rqthree}{\textbf{RQ1:}\textsl{How expert behaviour can be characterised? (expectations: they should be less biased to the position)}}

\renewcommand{\rqmain}{\textsl{How does technical domain expertise influence search behavior?}}
\renewcommand{\rqone}{\textbf{RQ1:} \textsl{How can we measure technical domain expertise?}}
\renewcommand{\rqtwo}{\textbf{RQ2:} \textsl{What is the impact of technical domain expertise on the search process?}}
\renewcommand{\rqthree}{\textbf{RQ3:} \textsl{What is the impact of technical domain expertise on search outcome?}}


\numberofauthors{1} 

\author{
\alignauthor 
Julia Kiseleva\(^1\)
$\qquad$ 
Alejandro Montes Garc{\'\i}a\(^1\)
$\qquad$ 
Jaap Kamps\(^2\)
$\qquad$ 
Nikita Spirin\(^3\)
\sharedaffiliation
\affaddr{\mbox{}\(^1\)Eindhoven University of Technology, Eindhoven, The Netherlands}\\
\affaddr{\normalsize \texttt{\{j.kiseleva,a.montes.garcia\}@tue.nl}}\\
\affaddr{\mbox{}\(^2\)University of Amsterdam, Amsterdam, The Netherlands}\\
\affaddr{\normalsize \texttt{kamps@uva.nl}}\\
\affaddr{\mbox{}\(^3\)University of Illinois at Urbana-Champaign, Urbana, IL}\\
\affaddr{\normalsize \texttt{spirin2@illinois.edu}}\\
}


\maketitle

\begin{abstract}
Domain expertise is regarded as one of the key factors impacting search success: experts are known to write more effective queries, to select the right results on the result page, and to find answers satisfying their information needs.
Search transaction 
logs play the crucial role in the result ranking. Yet despite the variety in expertise levels of users, all prior interactions are treated alike, suggesting that weighting in expertise can improve the ranking for informational tasks.
The main aim of this paper is to investigate the impact of high levels of technical domain expertise on both search behavior and task outcome. We conduct an online user study with searchers proficient in programming languages. We focus on Java and Javascript, yet we believe that our study and results are applicable for other expertise-sensitive search tasks.
The main findings are three-fold: 
First, we constructed expertise tests that effectively measure technical domain expertise and correlate well with the self-reported expertise. 
Second, we showed that there is a clear position bias, but technical domain experts were less affected by position bias.
Third, we found that general expertise helped finding the correct answers, but the domain experts were more successful as they managed to detect better answers. 
Our work is using explicit tests to determine user expertise levels, which is an important step toward fully automatic detection of expertise levels based on interaction behavior. 
A deeper understanding of the impact of expertise on search behavior and task outcome can enable more effective use of expert behavior in search logs\:---\:essentially make everyone search as an expert.
\end{abstract}

\vspace{-10pt}
\small
\vspace{-10pt}
\normalsize

\section{Introduction}
\label{sec:intro}

Users exhibit remarkably different search behavior, due to various differences including domain expertise that can greatly influence their ability to carry out successful searches.  The broad motivation of our work is to investigate how we can make non-experts search like experts.  Users' domain expertise is not the same as search expertise~\citep{White_wsdm_2009}. It concerns knowledge of the topic of the information need and it does not regard knowledge of the search process.  So far the differences in search behavior between experts and non-experts have been examined from the perspective of: \textbf{(1)} a query construction process; \textbf{(2)} search strategies; \textbf{(3)} and search outcomes. 
%
%
%
%
%
%
%
%
%
In early work, \citet{marc:info93} found that general experts in computer science, business/economics, and law, searched more content driven, and used more technical query terms. 
Using massive logs, \citet{White_wsdm_2009} observed differences in source selection (hostnames, TLDs), engagement (click-through rates), and vocabulary usage, for users with general expertise in medicine, finance, law, and computer science.
In a user study, \citet{cole:know11} show that task and domain knowledge are beneficial to link selection in medical literature search. 
In related work, \citet{zhan:pred11} shows that we can predict the medical domain expertise level based on the variables in a user study.

In this paper, we focus on high levels of technical domain expertise, specifically experts proficient in two programming languages: Java and JavaScript, and are interested in the distinction with general programming expertise.   This is different from earlier work comparing general experts against non-experts, as we focus on comparing degrees of technical domain expertise amongst experts, trying to find out if high levels of proficiency make a difference.
%
We study interactions with the Search Engine Result Page (SERP) by users with advanced technical domain expertise. 
The scenario of users' interactions with a SERP is simple: a user runs a query $Q$ and the search engine  retrieves the results ranked based on a relevance score~\citep{Craswell_2008}. 
We want to understand if technical domain experts are able to find {`better'} answers to their queries by exploring the SERP.  The study is motivated by our own frustration when using web search engines and Q\&A sites for technical how to questions, where the top ranked answers often are not the best results.

Evaluation of relevance ranking has traditionally relied on explicit human judgments or editorial labels. However, human judgments are expensive and difficult to derive from a broad audience.  Moreover, it is difficult to simulate realistic information needs.  User clicks are known as a good approximation towards obtaining \emph{implicit user feedback}~\citep{Agichtein_sigir_2006_1}.
This relies on the basic assumption that users click on relevant results.  Therefore, leveraging click-through data has become a popular approach for evaluating and optimizing information retrieval systems~\citep{Brandt_2011}.
However, it is well known that user behavior on the SERP is biased.  Different types of biases are discovered including:
\textbf{(1)} a \emph{position bias}~\citep{Joachims_2002, Joachims_2005}, users have a tendency to click on the first positions;
\textbf{(2)} a \emph{snippet bias}~\citep{Yue_2010}, snippets influence user decisions;
\textbf{(3)} a \emph{domain bias}~\citep{Ieong_2012}, that shows that users are already familiar with Internet and they are influenced by the domain of the URL; 
\textbf{(4)} a \emph{beliefs bias}~\citep{White_2013}, users beliefs affect their search behaviour and their decision making.
Taking into consideration these biases suggests that not all clicks are equally useful for optimizing a ranking function. For one thing only successful clicks (satisfied or SAT clicks) should be taken into consideration.
Our expectation is that users with a high level of proficiency are less affected by these biases.

\citet{ageev_sigir_2011} proposed flexible and general informational search success model for in-depth analysis of search success which was tested based on game-like infrastructure for crowdsourcing search behaviour studies, specifically targeted towards capturing and evaluating successful search strategies on informational tasks with known intent.

The main research question studied in this paper is: \rqmain\ 
We are particularly interested in what level of expertise is needed to make a difference: can we only trust the interactions of technical domain experts that essentially know the answers, or is a general familiarity with the domain sufficient? 
We conduct an user study with explicit tests to derive user expertise level, to determine the impact of technical domain expertise.  This is an important step toward fully automatic detection of expertise levels based on interaction behavior.
We have three concrete research questions:
\begin{description} \itemsep 0pt
\item \rqone\
\item \rqtwo\
\item \rqthree\
\end{description}

In order to answer our research questions, we organize a user study where we are trying to imitate a realistic scenario of search tasks in a technical domain.  Study participants are all having some programming background but different levels of knowledge of
two programming languages: Java and JavaScript.
A typical use case is that experts use search engines to re-find an answer which is common practise especially in the programming domain. For domain experts it is easy to verify if a page contains the required answer.  Participants with only general domain expertise typically are proficient in one programming language, say Java, but search for information in a unfamiliar programming language, say JavaScript.
Their fragmented understanding of the new language makes it much harder for them to recognize if a page contains the needed answers.
We imitate this scenario in our user study.




This paper is structured as follows:
\S\ref{sec:study_desc} details the setup of our user study, while 
\S\ref{sec:res} discusses the results, 
and we conclude in \S\ref{sec:conc}.

\shrink
\section{User Study Design}
\label{sec:study_desc}
\input{study_desc}

\shrink
\section{Results}
\label{sec:res}
\input{results}


\shrink
\section{Conclusions and Discussion}
\label{sec:conc}

The main aim of this paper was to investigate how technical domain expertise influences search behavior, focusing on high levels of proficiency in programming languages versus general programming expertise.  
We studied three concrete research questions.
First, we investigated: \rqone\
Our main finding was that the expertise test is effective and correlates well with the self-reported expertise.  
Second, we looked at: \rqtwo\
Our main finding was that the distribution of SAT clicks exhibited an evidence all participants were biased the URL position on SERP. However, the technical domain expert's biases was less pronounced and they tended to check the SERP's bottom.
Third, we examined: \rqthree\  
Our main finding was that having a general programming expertise helped to derive the good answers on the SERP, but the experts with high proficiency managed to detect better answers as they dug them from the bottom of the SERP. 

Our general conclusion is that participants with technical domain expertise behaved differently, and were more effective, than those with general expertise in the area.  The differences are clear, but mostly a matter of degree, suggesting that there is value in both types of interactions.  This suggests that properly weighting clicks relative to the expertise\:---\:and essentially use the expert behavior to get clicks of higher quality that hold the potential to improve the search result ranking.  
We are currently working on the prediction of technical domain expertise levels based on behavioral data.

Our results are on technical domain expertise, where levels of proficiency can be crisply defined, and we focused on searches related to their work task using web search engines and Q\&A sites for technical how to questions.  We expect our results to generalize to other specialized areas, typical of domain-specific search.  In light of the earlier literature on domain expertise, these results suggest that we need to go beyond the separation of those with and without domain expertise or familiarity\:---\:the classic distinction between experts and novices\:---\:but that there is value in distinguishing high levels of domain expertise\:---\:the distinction between general expertise in the area, and those `who know the answer.'


\subsection*{Acknowledgments}
The questionnaires, collected data, and the code for running the study, are available from \url{http://www.win.tue.nl/~mpechen/projects/capa/#Datasets}.

This research has been partly supported by STW and it is the part of the CAPA\footnote{www.win.tue.nl/$\sim$mpechen/projects/capa/} project.

\shrink
\renewcommand{\bibsection}{\section*{References}}
\bibliographystyle{abbrvnat}
\renewcommand{\bibfont}{\small}
\setlength{\bibhang}{1em}
\setlength{\bibsep}{0 pt}  

\balancecolumns

\end{document}

%% file: study_desc.tex
In this section, we explain the experimental setup of the online user study.

\mypar{Selecting Participants}
The call for participation was targeted to people interested in programming to make sure that they are likely to have information needs used in the study, using a snowball sampling approach.
By doing so, we tried to make our study as realistic as possible.
As an expertise field we selected two programming topics, namely, Java and JavaScript. However, users who are familiar with programming in general but not with the two selected topics also participated.  We tried to keep balance between the technical domain experts and the general experts in our study.

\mypar{Measuring Domain Expertise} 
Prior to starting the study, users were asked to \textbf{(1)} provide some basic demographic data, namely age, gender and education level; \textbf{(2)} self-report their programming level in general, but also their skills in Java and JavaScript.
Our study consist of two sets of tasks for each of the programming languages. The order of the topics in the study is done randomly.

Substantial research has been done in order to propose strategies to estimate users' expertise~\citep{White_wsdm_2009}. In order to 
have a reliable measurement, 
we ask participants to fill out a questionnaire related to one of the topics.
Users were not allowed to use a search engine in this step, they had to answer the questions based on their knowledge. Each questionnaire consists of ten questions.
We use the results of the questionnaire in order to identify a users' expertise level by assigning them a score $\in [0,1]$ based on their answers.

\mypar{Designing User Study}
%
\begin{figure*}[!t]
\includegraphics[width=\linewidth]{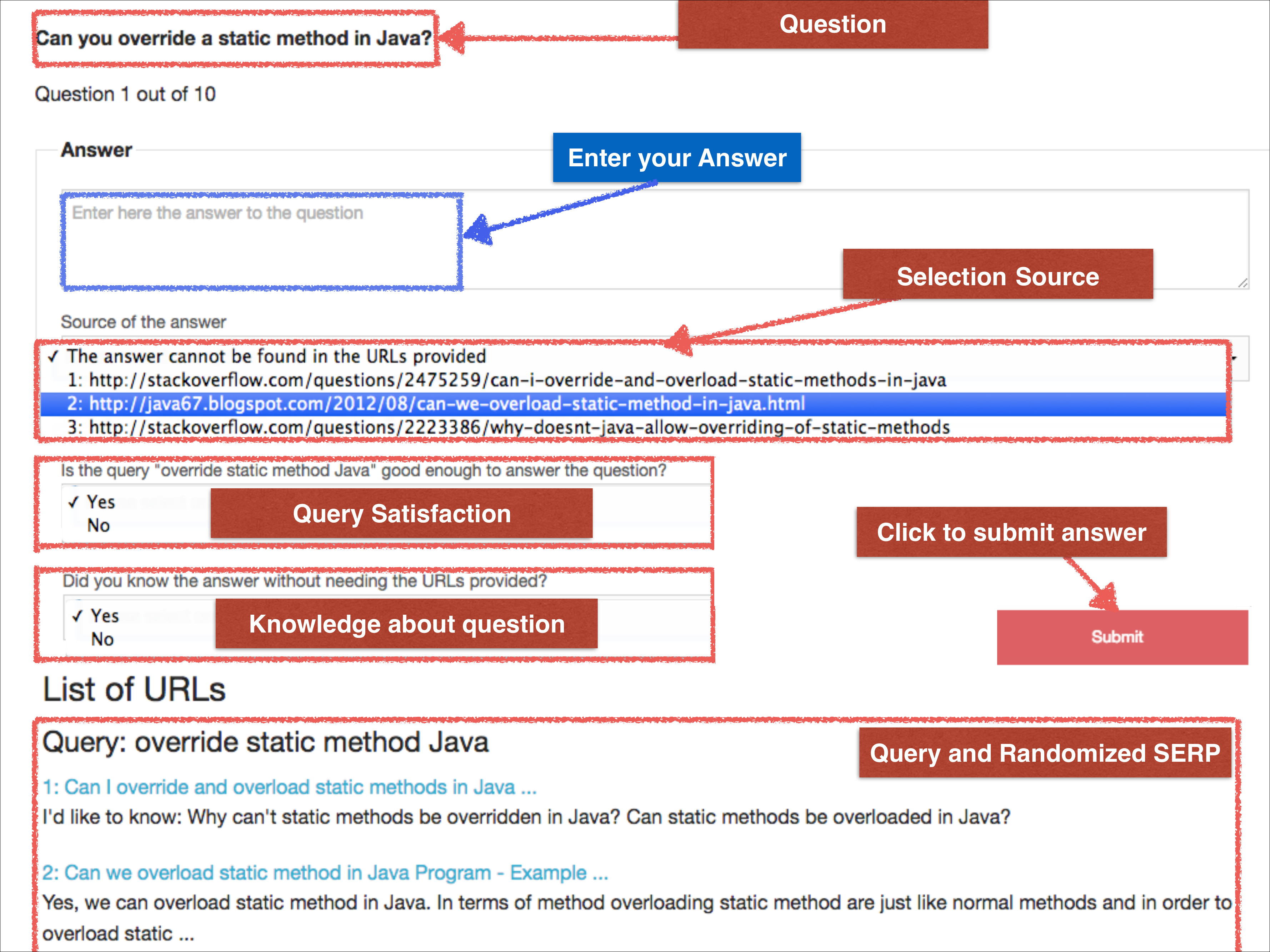}
\vspace{-10pt}
\caption{Screenshot of the online user study.}
\label{fig:screenshot}
\end{figure*}
%
The study consists of two sets of ten tasks or questions that related to the programming language.  Our tasks were modeled after those that users post in specialized Q\&A sites. For example, one of the tasks for Java was: ``Can you override a static method in Java?''
To illustrate the user study design, we provide a screenshot of one of the questions that the user had to fill, which can be seen in Figure~\ref{fig:screenshot}.

We provide a search query for each question that is used to retrieve a SERP with ten results from the Bing API.\footnote{\url{http://datamarket.azure.com/dataset/bing/search}}
The given SERP is randomly re-ranked in order to estimate the position bias.
The participants are asked to find the answer 
and to submit this URL. In addition, we ask participants \textbf{(1)} to tell us whether the query is formed in a right way or not; \textbf{(2)} to indicate if they knew the answer upfront. 
We collected ground truth answers from an expert who judged all shown results.



After finishing the study, users are asked to report again their proficiency in programming, Java and JavaScript to see if they changed their self-consideration after the test. Interestingly, five attenders have changed their mind. They could also enter open comments. For example, we have got the following comment: \emph{`Turns out there are some concepts which have `faded' a bit in my memory!'}, which basically shows that even good developers sometimes need to refresh their memory.



%% file: results.tex
In this section, we try to answer our three main research questions.
First, we look at the collected data and the results from the domain knowledge tests.
Second, we look at how the position bias depends on user expertise and how the number of correct answers depends on user expertise.


\mypar{Measuring Technical Domain Expertise}
We now describe the collected data, and try to answer our first research question: \rqone\ 

In total, we have 29 participants in our study. From the demographic perspective our dataset can be characterized by participants age and education level. In terms of education level we have the following population of participants: High school 8\%, Bachelor 12\%, Master 56\%, PhD 24\%. In terms of age we have the following population of participants: 18-23 years is 16\%, 24-29 years is 36\%, 30-35 years is 28\%, 36-42 years is 16\%, 43-48 years is 4\%.

\begin{figure}[!t]
\centerline{\includegraphics[width=1.0\linewidth]{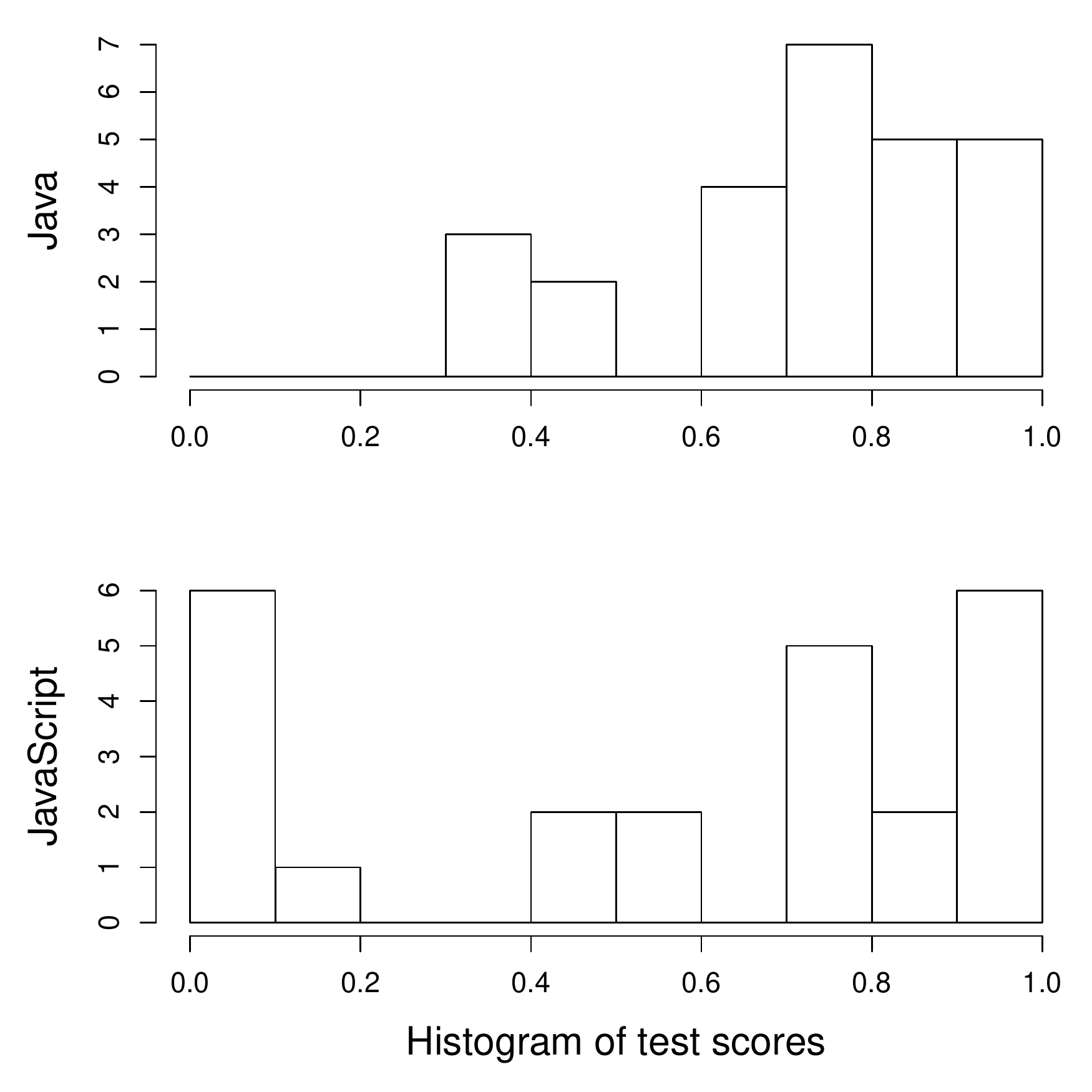}}
\vspace{-10pt}
\caption{A histogram of expertise score of participants calculated for Java (top) and JavaScript (bottom).}
\label{fig:hist_skill_score}
\end{figure}

Figure~\ref{fig:hist_skill_score} shows the distribution of expertise in Java (top half) and JavaScript (bottom half).
The expertise scores are calculated based on ten questions from the pre-survey (described in \S\ref{sec:study_desc}).
As we can see, the majority of the participants in the study have high levels of expertise in either Java or JavaScript.

\begin{figure}[!t]
\centerline{\includegraphics[width=1.0\linewidth]{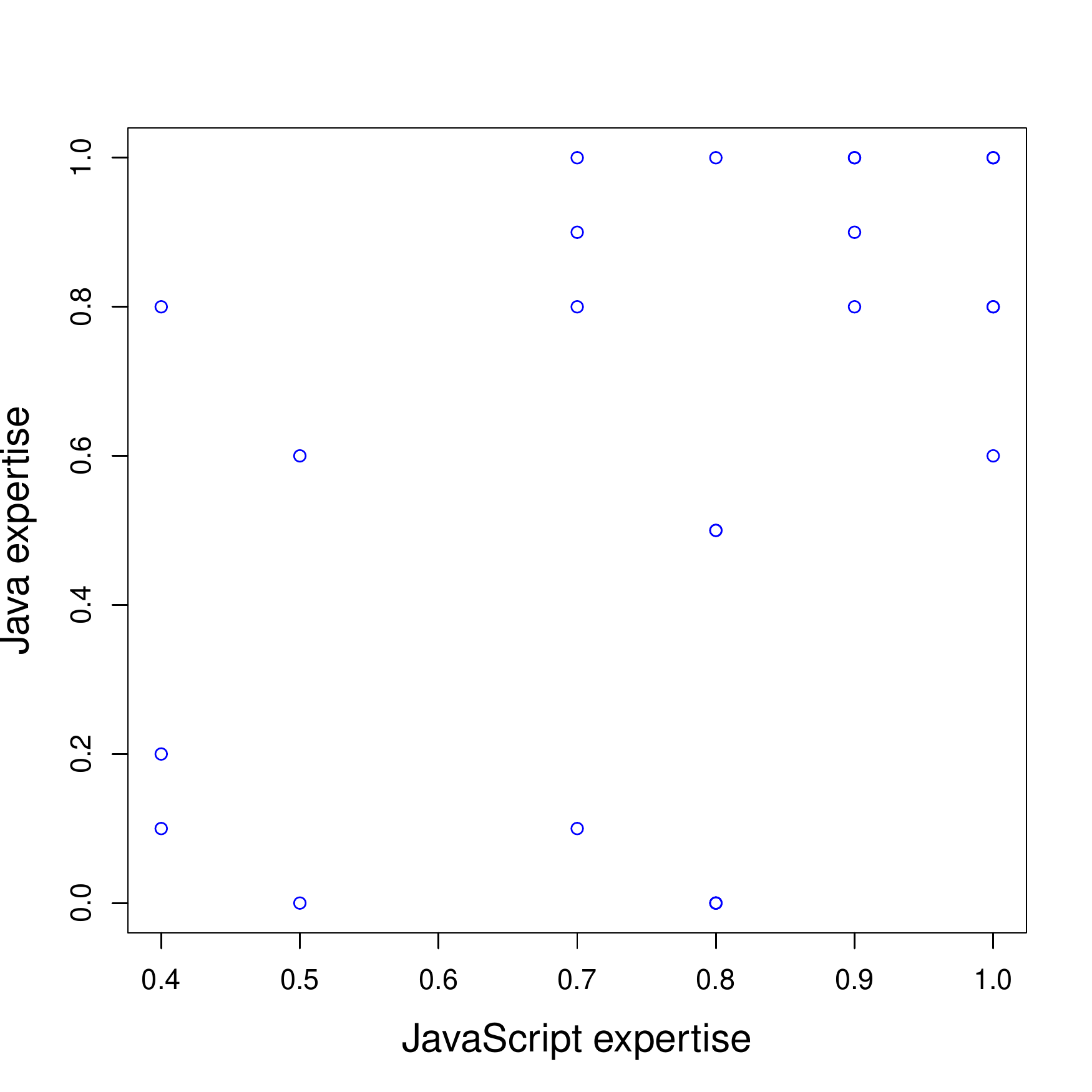}}
\vspace{-10pt}
\caption{A scatter plot of the calculated skill score for both topics: Java and JavaScript.}
\label{fig:skill_score}
\end{figure}

Figure~\ref{fig:skill_score} shows the relation between the participants' expertise scores in Java and JavaScript.
The relation between the skills in Java and JavaScript is weak (Pearson correlation of 0.44, $p < 0.05$) signaling that only a small fraction of participants has high levels of expertise in both.


\begin{figure*}[!t]

\centerline{
\includegraphics[width=0.45\linewidth]{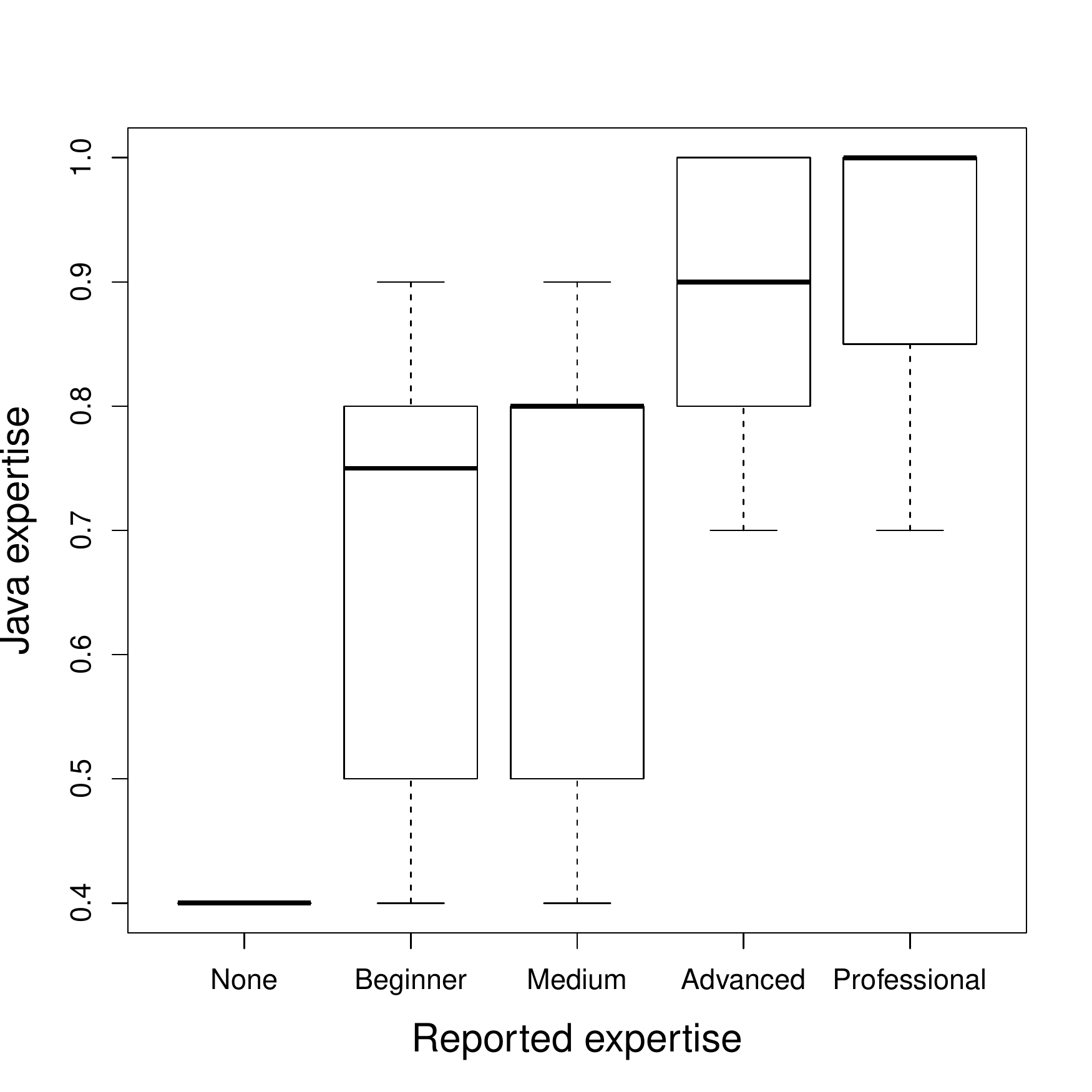}
~
\includegraphics[width=0.45\linewidth]{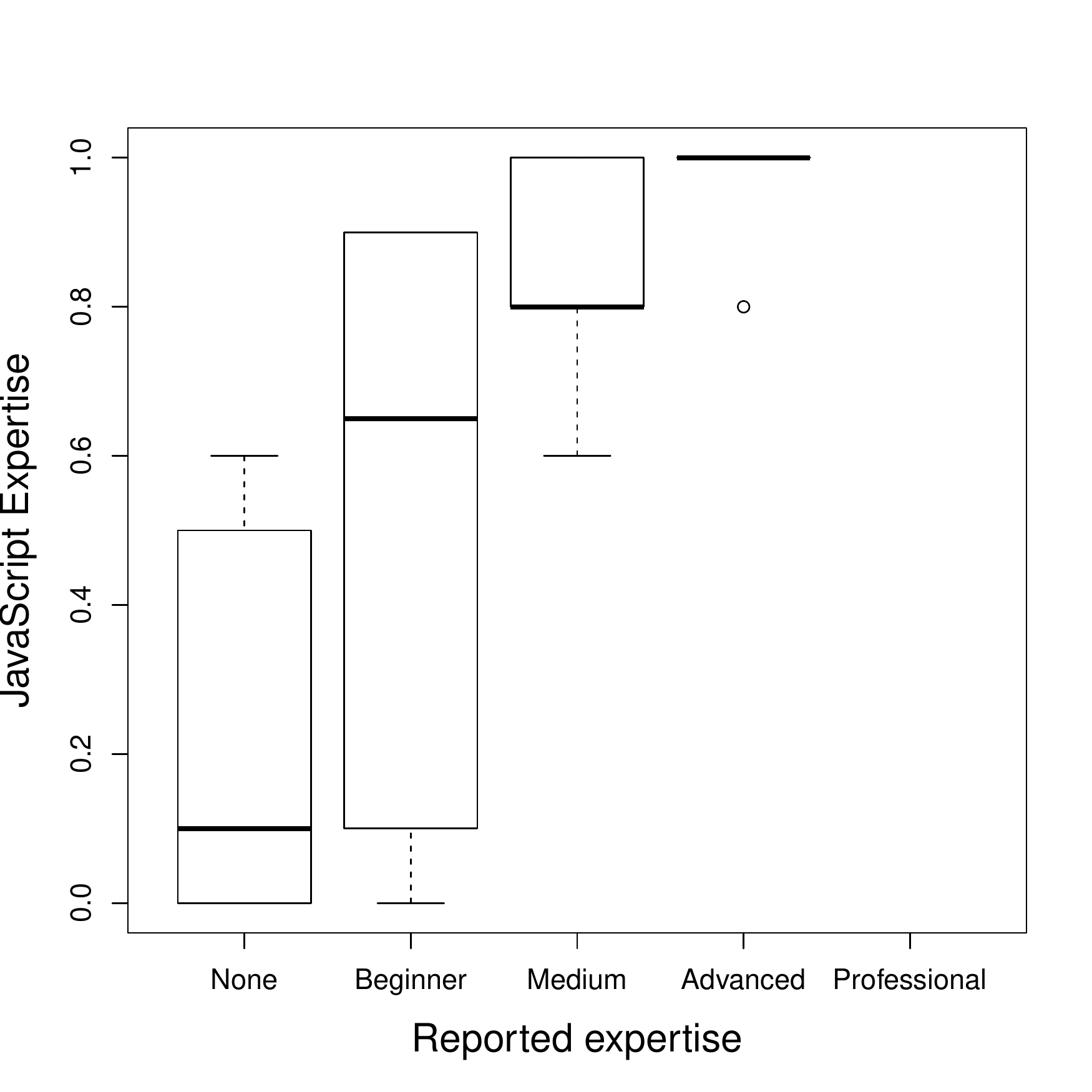}
}
\vspace{-10pt}
\caption{Expertise test results over self-reported scores for Java expertise (left) and JavaScript expertise (right).}
\label{fig:test_exp}
\end{figure*}
Figure~\ref{fig:test_exp} shows the expertise test scores over the self-reported expertise levels.  
We see a clear relation between the self reported expertise levels and the test scores, the correlation is 0.60 for Java (Pearson, $p < 0.01$)  and 0.75 for JavaScript ($p < 0.001$).  This result gives confidence in the tests to quantify the technical domain expertise of the participants.

\smallskip
Our main finding is that the expertise test is effective and correlates well with the self-reported expertise.  This implies the usefulness of the test, but also validates the self-reported expertise score as a reliable indicator of user expertise.

\mypar{Impact on Search Behavior}
We now investigate our second research question: \rqtwo\

\begin{figure*}[!t]
\centering
\includegraphics[width=0.65\linewidth]{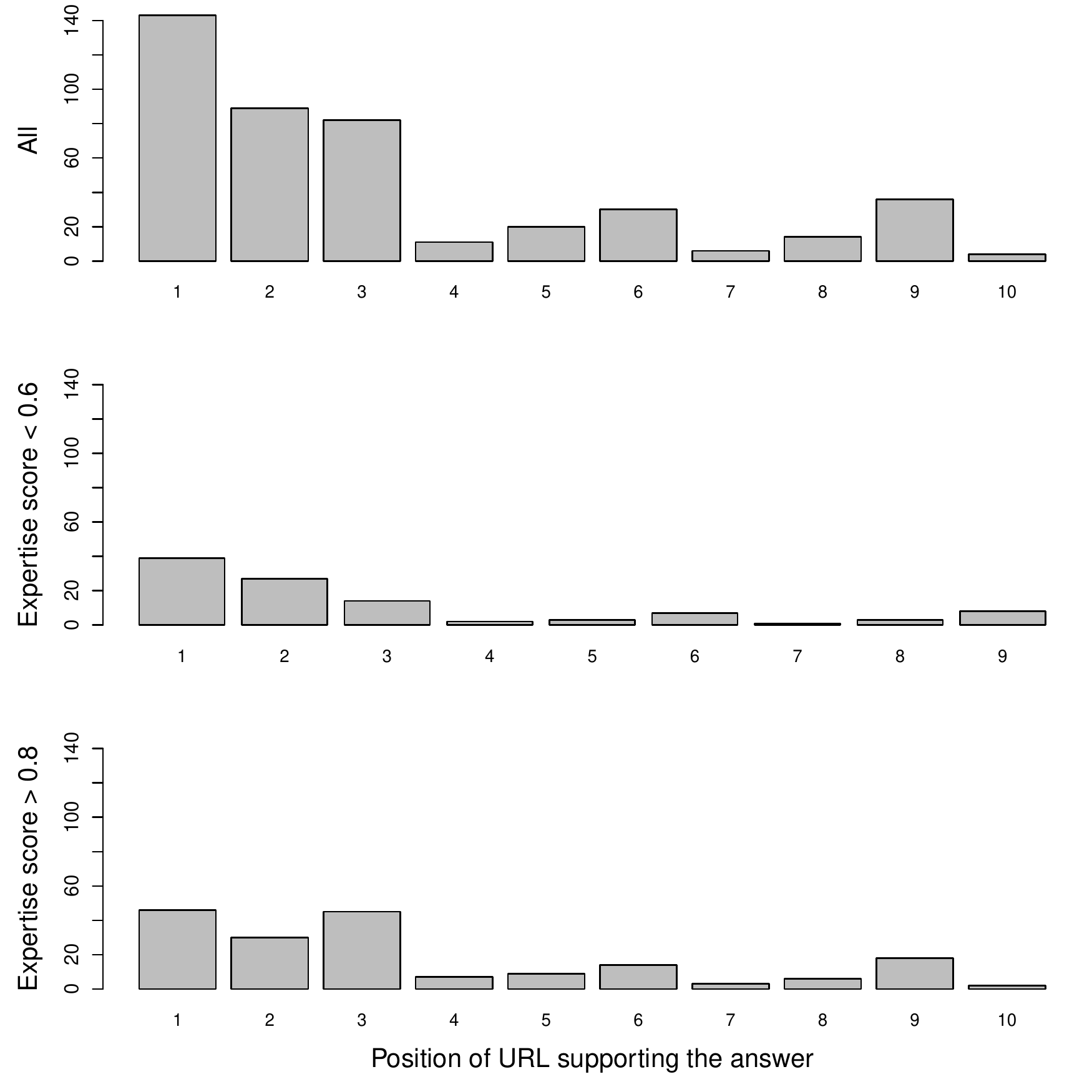}
\vspace{-10pt}
\caption{Position bias with respect to the users' expertise.}
\label{fig:pos_bias}
\end{figure*}

We calculate the distribution of positions over all submitted users answers for both topics (a position of URL that study participants submit as a source of the correct answer to the proposed question). This event can be mapped to the SAT click in web search behavior. 
Figure~\ref{fig:pos_bias} presents the distribution of selected answers with regard to the calculated users' expertise score: over all participants (top), over those with relatively low levels of expertise ($<$ 0.6, middle), and over those with a relatively high levels of expertise ($>$ 0.8, bottom). We performed a goodness of fit test against a uniform distribution which fails convincingly ($\chi^2$ goodness of fit test, $p < 0.001$ for all three cases).

Recall that the SERPs were randomized, hence relevant results are uniformly distributed across the ten positions.  We can clearly see that the whole population is biased to the position of the result on the SERP, showcasing that results are scanned from top to bottom.
We see an even stronger position bias for those with lower test scores, while for those with a higher test score, the position bias is less pronounced.  The technical experts more frequently pick a lower ranked results -- although even the experts clearly prefer top ranked results.  

The position bias as shown in Figure~\ref{fig:pos_bias} does not show a monotonically declining pattern, as some positions such as 6 and 9 are more popular than others.  Closer inspection reveals that this is due to the popularity particular Q\&A sites, in particular \url{http://stackoverflow.com/}, that attract attention.  As we are working with a single randomized SERP for each question\:---\:allowing us to compare position across participants\:---\:the distribution of popular Q\&A sites is not exactly uniform over the sample. So in addition to the position bias, we see a domain bias.

We can see an evidence of the snippet bias in participants behavior as they are selecting `correct' result URLs without clicking on them.  Indeed, for these cases we can see that answers are provided in snippets.

%



\smallskip
Our main finding is that the click distribution shows evidence of the position bias of all participants. However, for the experts the position bias is less pronounced and they tend to check the lower positioned results.

\mypar{Impact on Search Outcome}
We now examine our third research question: \rqthree\

In order to collect the ground truth for correctness of the provided answers, we pooled results from the higher scoring experts and had the answers and URLs judged by an expert editorial judge.  
As it turned out, on average 5 out of 10 results supported answering the task, varying between 2 and 8 per question.

\begin{figure*}[!t]
\centerline{
\includegraphics[width=0.45\linewidth]{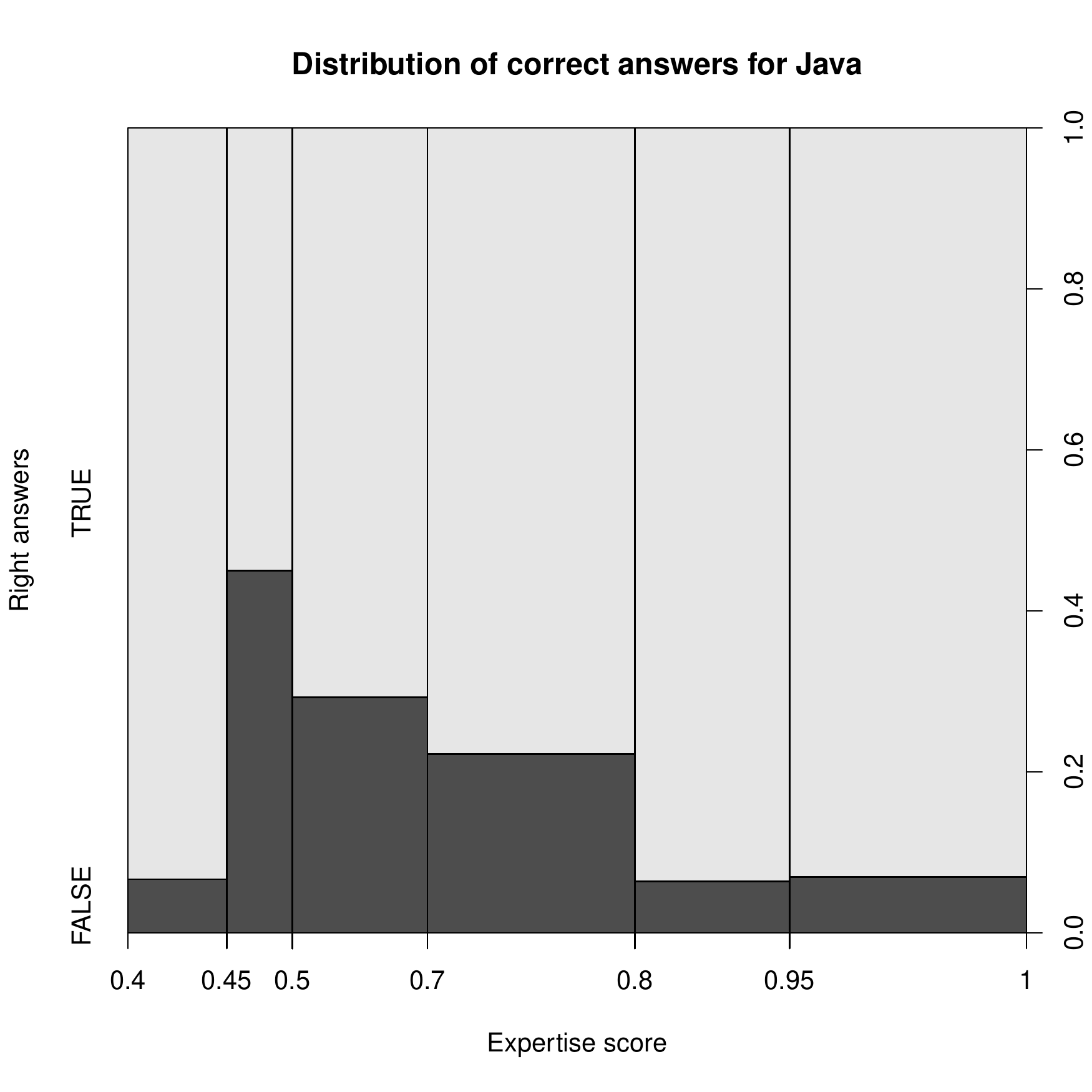}
~
\includegraphics[width=0.45\linewidth]{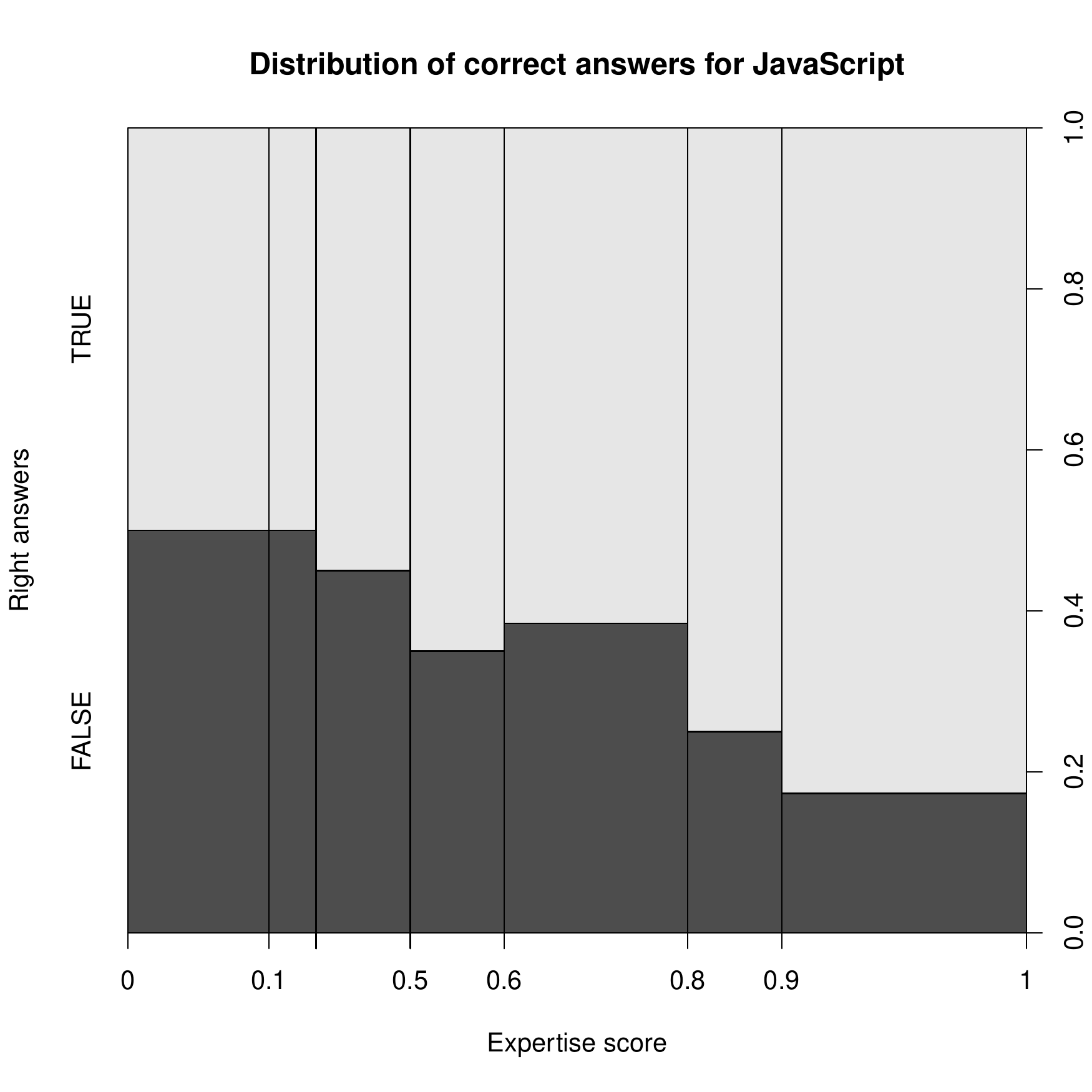}}
\vspace{-10pt}
\caption{A distribution of right answers of participants with different expertise levels in Java (left) and JavaScript (right).}
\label{fig:answers}
\end{figure*}

Figure~\ref{fig:answers} shows the distribution of correct answers over expertise levels for Java (left) and for JavaScript (right).  We see a clear relation for both Java and Java script: higher expertise levels lead to higher fractions of correct answers.  The relation is highly significant (Pearson $\chi^2$, $p<0.0001$) for both Java and JavaScript.
There is an interesting deviation for those scoring very low on Java, yet producing many correct answers. A plausible explanation it that these participants have sufficient passive understanding, hence can recognize answers pages on the information on the web pages, but cannot actively produce this in the test.  This is supported by their relatively high fractions of ``I don't know" answers on the Java expertise test.

\smallskip
Our main finding is that the participants with general programming expertise are able to find out the correct answers on SERP but they tend to select higher positioned pages in SERP. Clearly, the experts manage to detect better answers as they dig them from the bottom of SERP. 